\documentclass[11pt,a4paper]{article}
\usepackage[utf8]{inputenc}
\usepackage{geometry}
\usepackage{amsmath,amssymb,amsfonts}
\usepackage{graphicx}
\usepackage{hyperref}
\usepackage{float}
\usepackage{subcaption}
\usepackage{tabularx}
\usepackage{cite}
\usepackage{color}
\usepackage{authblk}

\hypersetup{
    colorlinks=true,
    linkcolor=blue,
    filecolor=magenta,
    urlcolor=blue,
    citecolor=red,
}

\title{\textbf{Cosmological Constraints on 4D Einstein-Gauss-Bonnet Gravity and Kaniadakis Holographic Dark Energy: Implications for Black Hole Shadows}}

\author[1]{Xiang-Qian Li\thanks{Corresponding author: lixiangqian@tyut.edu.cn}}
\author[1]{Hao-Peng Yan}
\author[1]{Xiao-Jun Yue}
\affil[1]{College of Physics and Optoelectronic Engineering, Taiyuan University of Technology, Taiyuan 030024, China}

\date{\today}

\begin{document}

\maketitle

\begin{abstract}
The direct imaging of black holes by the Event Horizon Telescope (EHT) enables strong-field tests of gravity. We study the cosmological evolution and the black-hole shadow radius in 4D Einstein-Gauss-Bonnet (EGB) gravity coupled to Kaniadakis holographic dark energy (KHDE), adopting the future event horizon as the infrared cutoff. Using Cosmic Chronometers, Pantheon+ Type Ia supernovae, and DESI BAO data, we constrain the model with a Markov Chain Monte Carlo analysis. The best-fit values favor a phantom-like equation of state driven by Kaniadakis entropy ($c\simeq 1.18$, $\beta\simeq 2.26$), but $\beta$ remains weakly constrained ($\beta=2.26^{+0.11}_{-2.20}$), consistent with the standard holographic limit $\beta\to0$ at $1\sigma$. The EGB coupling is constrained to $\alpha\simeq -0.004$, also consistent with General Relativity ($\alpha=0$) at $1\sigma$. Guided by the posterior, we define five representative scenarios to probe the dynamical phase space. We find that the accretion history is highly sensitive to the thermodynamic sector: standard holographic cases yield monotonic evolution, whereas phantom-divide crossing leads to non-monotonic behavior in both the black hole mass and the vacuum shadow radius. Including a dispersive plasma medium, refraction dominates over intrinsic mass growth and induces an overall shrinkage of the observable shadow at high redshift; nevertheless, a residual intrinsic deviation of $\sim6\%$ (for our conservative accretion setup) persists at $z\simeq2$ relative to the $\Lambda$CDM prediction. These results indicate that, despite environmental dominance, precision population analyses of black hole shadows may help disentangle subtle dynamical dark-energy imprints from the standard cosmological paradigm.
\end{abstract}

\tableofcontents
\vspace{1cm}

\section{Introduction}
\label{sec:intro}

The advent of the Event Horizon Telescope (EHT) has ushered in a precision era for testing fundamental physics in the strong-field regime. Horizon-scale images of the supermassive black holes M87* \cite{EHT2019} and Sagittarius A* (Sgr A*) \cite{EHT2022} provide a unique avenue to probe the underlying spacetime geometry through the ``black hole shadow.'' Although current results are consistent with General Relativity (GR), the present uncertainties still allow viable departures from GR. This remains particularly timely in view of the open challenges faced by GR on cosmological scales, most notably the origin of late-time acceleration and the initial singularity problem.

Motivated by these issues, theoretical efforts commonly extend the standard framework by modifying either the gravitational sector or the energy--momentum content. On the gravity side, the 4D Einstein-Gauss-Bonnet (EGB) theory proposed by Glavan and Lin \cite{Glavan2020} has attracted substantial attention. By rescaling the Gauss--Bonnet coupling as $\alpha \to \alpha/(D-4)$ and taking the limit $D\to4$, the theory yields non-trivial black-hole solutions that formally evade the Lovelock theorem. While the original regularization procedure has been questioned in the literature \cite{Lu2020, Ai2020}, subsequent well-defined scalar--tensor formulations \cite{Hennigar2020, Fernandes2020} have clarified the status of the resulting field equations for static, spherically symmetric spacetimes. Since then, a broad body of work has explored phenomenological consequences of 4D EGB gravity, including black-hole thermodynamics \cite{Wei2020, Hegde2020, Halder:2025lrq, Hu:2025mdr}, quasinormal modes \cite{Churilova2020, Konoplya2020a, Becar:2025niq}, and in particular shadow-related observables \cite{Zubair:2023cep, Badia:2021kig, Kumar2020, Liu:2022plm, Vagnozzi:2022moj}. On cosmological scales, the theory admits negative coupling ($\alpha<0$), which can support non-singular bouncing solutions \cite{Khodabakhshi2024} and has been constrained by Big Bang Nucleosynthesis and late-time expansion data \cite{Zanoletti:2023ori, Clifton2020}.

In parallel, the dark-energy sector has been enriched by ideas from non-extensive statistical mechanics. Kaniadakis ($\kappa$-) statistics \cite{Kaniadakis2002, Kaniadakis2005} provides a generalized entropy that reduces to the Boltzmann--Gibbs form in the limit $\kappa\to0$. More broadly, Kaniadakis statistics is part of a wider class of generalized entropies relevant for holographic cosmology, as emphasized in recent classifications \cite{Nojiri:2022dkr}. In this setting, the Kaniadakis holographic dark energy (KHDE) model adopted here can be viewed as a specific realization within the generalized holographic dark energy framework originally introduced by Nojiri and Odintsov \cite{Nojiri:2005pu}. While such generalized entropic constructions offer a versatile phenomenological arena, it has been noted that black holes endowed with non-extensive entropies may encounter thermodynamic stability issues \cite{Nojiri:2022aof, Elizalde:2025iku}. Nevertheless, KHDE has shown promise in alleviating the $H_0$ tension \cite{Hernandez2022} and can accommodate phantom-divide crossing ($w<-1$) without introducing exotic matter fields \cite{Shen2025cjm, Lymperis2021}.

A key missing link, however, is the connection between global cosmological dynamics and local, horizon-scale observables. Many studies of shadows in modified gravity assume a stationary black-hole mass and a vacuum propagation medium, thereby neglecting potential couplings to the evolving cosmic background. In principle, a black hole embedded in a cosmological fluid can experience a secular mass drift; within the Babichev accretion prescription \cite{Babichev2004}, phantom-like dark energy can even drive mass loss, in contrast to standard accretion. Moreover, realistic observations are performed through dispersive plasma, which can modify photon trajectories via refraction and hence alter the inferred shadow size \cite{Perlick2015, Xu:2025iwg}. Since the plasma distribution near supermassive black holes is environment dependent and not expected to universally track the cosmic mean, one may adopt phenomenological scaling ans\"atze to quantify possible redshift trends and assess parameter degeneracies.

In this work, we develop a framework that connects cosmological constraints to the redshift dependence of the shadow radius of 4D EGB black holes in a KHDE background. Our analysis proceeds in three steps. First, we perform a Markov Chain Monte Carlo (MCMC) analysis using Cosmic Chronometers (CC), Type Ia Supernovae (SNIa), and Baryon Acoustic Oscillations (BAO) measurements from DESI to constrain the parameter set $(H_0,\Omega_{m0},\beta,\alpha,\ln c,r_{drag})$ without imposing a $\Lambda$CDM background. Second, based on these constraints, we model the secular mass evolution of supermassive black holes sourced by KHDE accretion, treating the accretion efficiency as a phenomenological parameter. Third, we compute the redshift dependence of the shadow radius by incorporating both the metric corrections from 4D EGB gravity and dispersive plasma effects along the photon path. Our aim is to determine whether intrinsic signatures of modified gravity and dynamical dark energy can survive, at least statistically, beneath the dominant environmental imprint of plasma refraction.

The remainder of this paper is organized as follows. Section~\ref{sec:theory} summarizes the theoretical setup. Section~\ref{sec:constraints} presents the observational constraints. Sections~\ref{sec:mass} and \ref{sec:shadow} analyze the black-hole mass evolution and the shadow radius, respectively. Finally, Section~\ref{sec:conclusion} concludes.

\section{Theoretical Framework}
\label{sec:theory}

This section summarizes the theoretical setup, namely 4D Einstein-Gauss-Bonnet (EGB) gravity coupled to Kaniadakis holographic dark energy (KHDE) with the future event horizon as the infrared cutoff. We first present the regularized 4D EGB black-hole geometry, then introduce the KHDE density and the modified Friedmann dynamics, and finally specify the secular black-hole mass evolution induced by dark-energy accretion in the inferred cosmological background.

\subsection{4D Einstein-Gauss-Bonnet Gravity}

The Einstein-Gauss-Bonnet action in $D$ dimensions can be written as
\begin{equation}
    S = \int d^Dx \sqrt{-g} \left[ \frac{R}{16\pi G} + \alpha_{GB} \mathcal{G} + \mathcal{L}_{m} \right],
\end{equation}
where $R$ is the Ricci scalar, $\mathcal{L}_m$ is the matter Lagrangian, and
$\mathcal{G} = R^2 - 4R_{\mu\nu}R^{\mu\nu} + R_{\mu\nu\rho\sigma}R^{\mu\nu\rho\sigma}$
is the Gauss--Bonnet invariant. Although $\mathcal{G}$ is topological in $D=4$, Glavan and Lin \cite{Glavan2020} proposed a regularization based on the rescaling $\alpha_{GB}\to \alpha/(D-4)$ and taking the limit $D\to4$ at the level of the field equations, leading to non-trivial 4D dynamics.

We consider a static, spherically symmetric ansatz
$ds^2 = -f(r)dt^2 + f(r)^{-1}dr^2 + r^2 d\Omega^2$.
Solving the vacuum field equations yields \cite{Glavan2020, Fernandes2022}
\begin{equation}
    f(r) = 1 + \frac{r^2}{2\alpha} \left( 1 - \sqrt{1 + \frac{8\alpha M}{r^3}} \right),
    \label{eq:metric_f}
\end{equation}
where $M$ is the geometric mass and $\alpha$ is the (rescaled) EGB coupling (we use geometric units $G=c=1$).

The coupling $\alpha$ controls the spacetime structure. The branch that is continuously connected to the Schwarzschild solution is recovered as $\alpha\to 0$, and expanding Eq.~(\ref{eq:metric_f}) reproduces the GR limit. For $\alpha<0$, reality of the metric requires
$1 + 8\alpha M/r^3 \ge 0$.
In the following, Eq.~(\ref{eq:metric_f}) provides the background geometry used in our shadow-radius computation.

\subsection{KHDE with Future Event Horizon Cutoff}

The holographic principle suggests that an effective vacuum energy density is constrained by an infrared (IR) cutoff scale. We adopt the Kaniadakis entropy (motivated by relativistic statistical mechanics), which can be written as $S_{\kappa} = \frac{1}{K} \sinh(K S_{BH})$ \cite{Hsu:2004ri}. In standard holographic dark energy, choosing the Hubble radius ($L=H^{-1}$) as the IR cutoff typically fails to produce late-time acceleration without additional interactions. We therefore identify the IR cutoff with the \textbf{future event horizon} $R_h$,
\begin{equation}
    R_h(t) = a(t) \int_t^{\infty} \frac{dt'}{a(t')} = a(t) \int_{a(t)}^{\infty} \frac{da'}{H(a') a'^2},
    \label{eq:future_horizon}
\end{equation}
which yields a phenomenologically viable accelerating phase.

For numerical convenience, we introduce the grouped parameter $\beta \equiv K^2 \pi^2 M_p^4$ to encode the non-extensive correction. The resulting KHDE density reads \cite{Drepanou2021, Luciano:2025ykr}
\begin{equation}
    \rho_{DE} = 3 c^2 M_p^2 \left( \frac{1}{R_h^2} + \frac{\beta R_h^2}{6} \right),
    \label{eq:rho_KHDE}
\end{equation}
where $c$ is a dimensionless constant and $M_p$ is the reduced Planck mass. In the extensive limit $\beta\to0$, Eq.~(\ref{eq:rho_KHDE}) reduces to the standard holographic form $\rho_{DE}=3c^2 M_p^2 R_h^{-2}$.

\subsection{Modified Friedmann Equations and Evolution}

We consider a spatially flat Friedmann--Robertson--Walker universe sourced by pressureless matter $\rho_m$ and KHDE $\rho_{DE}$. In 4D EGB cosmology, the Hubble rate $H\equiv \dot a/a$ satisfies \cite{Clifton2020}
\begin{equation}
    H^2 + \alpha H^4 = \frac{1}{3M_p^2}\,(\rho_m + \rho_{DE}).
    \label{eq:Friedmann_EGB}
\end{equation}
Defining $E(z)\equiv H(z)/H_0$ and $\tilde{\alpha}\equiv \alpha H_0^2$, the physical (GR-connected) branch can be expressed as
\begin{equation}
    E(z) = \left[ \frac{-1 + \sqrt{1 + 4\tilde{\alpha} \mathcal{R}(z)}}{2\tilde{\alpha}} \right]^{1/2},
    \label{eq:Ez_solution}
\end{equation}
where
\begin{equation}
    \mathcal{R}(z) \equiv \frac{\rho_{tot}(z)}{\rho_{crit,0}}
    = \Omega_{m0}(1+z)^3 + \frac{\rho_{DE}(z)}{3M_p^2 H_0^2},
    \qquad
    \rho_{crit,0}=3M_p^2 H_0^2.
    \label{eq:source_term}
\end{equation}
Here, $\rho_{DE}(z)$ is specified by Eq.~(\ref{eq:rho_KHDE}) through the future event horizon $R_h(z)$, and therefore depends non-locally on the expansion history.

Because $R_h$ is defined through an integral over $H$, the background evolution cannot be obtained from a purely algebraic relation. Differentiating Eq.~(\ref{eq:future_horizon}) and using $\dot R_h = H R_h - 1$, we obtain an ordinary differential equation in redshift,
\begin{equation}
    \frac{dR_h}{dz} = \frac{1}{1+z} \left( R_h(z) - \frac{1}{H(z)} \right).
    \label{eq:Rh_ode}
\end{equation}
In our MCMC analysis, we solve Eq.~(\ref{eq:Ez_solution}) together with Eq.~(\ref{eq:Rh_ode}) to obtain a self-consistent evolution of $H(z)$ and $R_h(z)$.

The dark-energy equation of state follows from the continuity equation
$\dot{\rho}_{DE} + 3H(1+w_{DE})\rho_{DE} = 0$.
Differentiating Eq.~(\ref{eq:rho_KHDE}) and using $\dot R_h = H R_h - 1$, we find
\begin{equation}
    w_{DE}(z) = -1 + \frac{2}{3} \left( \frac{\beta R_h^4(z) - 6}{\beta R_h^4(z) + 6} \right) \left( \frac{1}{H(z) R_h(z)} - 1 \right).
    \label{eq:wDE_z}
\end{equation}

We define the (time-dependent) dark-energy density parameter as
\begin{equation}
    \Omega_{DE}(z) \equiv \frac{\rho_{DE}(z)}{3M_p^2 H^2(z)}.
    \label{eq:Omega_def}
\end{equation}
This definition uses the usual critical-density normalization $3M_p^2H^2$ as a bookkeeping measure of the fractional contribution, even though the background dynamics are governed by the modified Friedmann constraint in Eq.~(\ref{eq:Friedmann_EGB}). In the extensive limit $\beta\to0$, Eq.~(\ref{eq:rho_KHDE}) implies
\begin{equation}
    \Omega_{DE}=\frac{c^2}{H^2R_h^2}
    \quad \Rightarrow \quad
    HR_h=\frac{c}{\sqrt{\Omega_{DE}}}.
\end{equation}
Substituting this relation into Eq.~(\ref{eq:wDE_z}) for $\beta\to0$ (for which the prefactor approaches $-1$), we obtain
\begin{equation}
    w_{DE}^{\beta \to 0} = -1 - \frac{2}{3} \left( \frac{1}{H R_h} - 1 \right)
    = -\frac{1}{3} - \frac{2\sqrt{\Omega_{DE}}}{3c},
\end{equation}
which reproduces the standard holographic dark energy result of Li \cite{Li:2004rb}. This limit provides a useful consistency check and shows that our formulation continuously connects to the standard holographic phenomenology.

\subsection{Black Hole Mass Accretion Dynamics}

We model the interaction between the black hole and the cosmological background through quasi-stationary, spherically symmetric accretion of an effective dark-energy fluid. In the Babichev prescription \cite{Babichev2004}, the mass change rate is
\begin{equation}
    \dot{M} = 4 \pi A M^2 (\rho_{DE} + p_{DE})
    = 4 \pi A M^2 \rho_{DE} (1 + w_{DE}),
\end{equation}
where $A$ is an effective accretion-efficiency parameter. Converting to redshift dependence gives
\begin{equation}
    \frac{dM}{dz} = - \frac{4\pi A M(z)^2 \rho_{DE}(z)\,[1 + w_{DE}(z)]}{H(z)(1+z)}.
    \label{eq:mass_acc_z}
\end{equation}
The sign of $(1+w_{DE})$ controls the secular drift: in the quintessence regime ($w_{DE}>-1$) the black-hole mass increases, whereas for phantom-like dark energy ($w_{DE}<-1$) it decreases with cosmic time.

In this work, $M(z)$ is interpreted as the secular mass drift sourced solely by dark-energy accretion within the Babichev model, rather than a complete description of SMBH growth. Standard astrophysical channels (e.g.~baryonic disc accretion, mergers, and feedback) are not included; their net impact is effectively absorbed into the choice of the reference mass at $z=0$ and/or into the phenomenological parameter $A$.

In hydrodynamic treatments, $A$ can be fixed by a critical (sonic) point analysis \cite{Mukherjee2024, Biswas2011}. In holographic dark energy models with phantom-divide crossing, however, the squared sound speed $c_s^2$ may develop divergences as $w\to -1$, making the usual sonic-point condition ill-defined. We therefore treat $A$ as a phenomenological constant and adopt $4\pi A=0.1$ to represent a conservative, low-efficiency secular accretion scenario. Equation~(\ref{eq:mass_acc_z}) is solved numerically using the background tracks $H(z)$ and $w_{DE}(z)$ inferred from the MCMC constraints.
\section{Observational Constraints}
\label{sec:constraints}

To assess the viability of the proposed scenario, we constrain the model parameters through a Bayesian analysis. We perform Markov Chain Monte Carlo (MCMC) sampling with the \texttt{Cobaya} package \cite{Torrado2021}, using a custom likelihood that implements the modified background evolution derived in Section~\ref{sec:theory}. The joint likelihood is taken as $\mathcal{L}\propto \exp(-\chi^2_{\rm tot}/2)$, where $\chi^2_{\rm tot}$ is the sum of the chi-square contributions from the individual datasets.

\subsection{Methodology and Datasets}

We sample the parameter vector $\Theta=\{H_0,\Omega_{m0},\alpha,\beta,\ln c,r_{drag}\}$ with flat priors. Our analysis is based on three complementary late-time geometric probes that do not rely on early-universe assumptions.

\paragraph{Cosmic Chronometers (CC):}
We use 31 model-independent measurements of the Hubble parameter $H(z)$ inferred from differential ages of passively evolving galaxies in the redshift range $0.07<z<1.965$ \cite{Moresco2016}. The corresponding chi-square is
\begin{equation}
    \chi_{CC}^2 = \sum_{i=1}^{31} \left( \frac{H_{\rm th}(z_i,\Theta)-H_{\rm obs}(z_i)}{\sigma_{CC}(z_i)} \right)^2,
\end{equation}
where $H_{\rm th}(z_i,\Theta)$ is the theoretical prediction and $H_{\rm obs}(z_i)$ with uncertainty $\sigma_{CC}(z_i)$ denotes the observation.

\paragraph{Type Ia Supernovae (SNIa):}
We employ the Pantheon+ compilation \cite{Brout2022}, consisting of 1701 Type Ia supernova light curves spanning $0.001<z<2.26$. The theoretical distance modulus is
\begin{equation}
    \mu_{\rm th}(z,\Theta)=5\log_{10}\!\left[(1+z)\int_{0}^{z}\frac{dz'}{E(z',\Theta)}\right]+25-5\log_{10}H_0.
\end{equation}
Because the absolute magnitude $M$ is degenerate with $H_0$ through the nuisance combination $\mathcal{M}=M-5\log_{10}H_0+25$, we analytically marginalize over $\mathcal{M}$ following the standard procedure \cite{SNLS:2011lii}. This allows SNIa to constrain the \emph{shape} of $E(z)$, while the absolute scale is informed by CC and BAO. The marginalized chi-square is
\begin{equation}
    \chi_{SNIa}^2 = \Delta \boldsymbol{\mu}^T \mathbf{C}^{-1} \Delta \boldsymbol{\mu}
    - \frac{\big(\Delta \boldsymbol{\mu}^T \mathbf{C}^{-1}\mathbf{1}\big)^2}{\mathbf{1}^T \mathbf{C}^{-1}\mathbf{1}},
\end{equation}
where $\Delta \boldsymbol{\mu}$ is the residual vector prior to applying $\mathcal{M}$, $\mathbf{C}$ is the full covariance matrix including systematics, and $\mathbf{1}$ is a vector of ones.

\paragraph{Baryon Acoustic Oscillations (BAO):}
We include the recent BAO measurements from DESI DR2 \cite{DESI:2025zpo}. Depending on the tracer sample, the observables are the transverse comoving distance ratio $D_M/r_{drag}$, the Hubble distance ratio $D_H/r_{drag}$, and/or the isotropic volume-distance ratio $D_V/r_{drag}$. The sound horizon at the drag epoch, $r_{drag}$, is treated as a free nuisance parameter. The BAO contribution is
\begin{equation}
    \chi_{BAO}^2 = \Delta \mathbf{V}_{BAO}^T \mathbf{C}_{BAO}^{-1} \Delta \mathbf{V}_{BAO},
\end{equation}
where $\Delta \mathbf{V}_{BAO}$ is the difference between theoretical predictions and the observed BAO data vector, and $\mathbf{C}_{BAO}$ is the corresponding covariance matrix.

Assuming these probes are independent, the total chi-square is
\begin{equation}
    \chi_{tot}^2 = \chi_{CC}^2 + \chi_{SNIa}^2 + \chi_{BAO}^2.
\end{equation}

\subsection{Results and Discussion}

\begin{figure}[ht]
    \centering
    \includegraphics[width=0.9\textwidth]{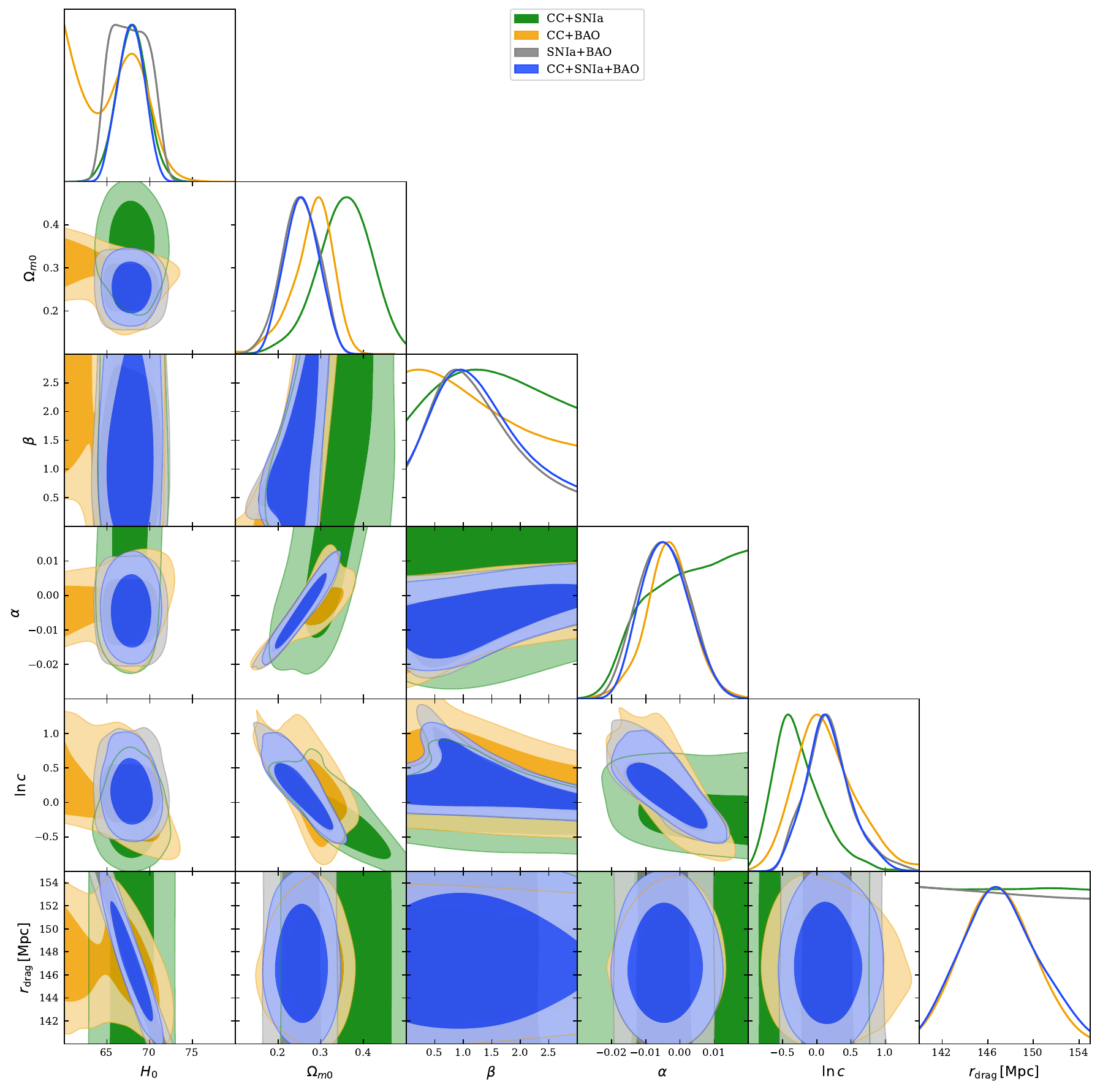}
    \caption{Confidence contours ($68\%$ and $95\%$) and 1D posterior distributions for the model parameters ($H_0, \Omega_{m0}, \beta, \alpha, \ln c, r_{drag}$) obtained from the joint analysis of CC+SNIa+BAO (blue), compared with partial combinations. The contours highlight the impact of including DESI BAO data on constraining $\beta$ and $\alpha$.}
    \label{fig:corner}
\end{figure}

We run MCMC chains for four dataset combinations: CC+SNIa, CC+BAO, SNIa+BAO, and the full combination CC+SNIa+BAO. The joint posterior distributions and 2D confidence contours for the full combination are shown in Fig.~\ref{fig:corner}, and the marginalized constraints for all combinations are summarized in Table~\ref{tab:constraints}.

\begin{table*}[tbp]
\centering
\renewcommand{\arraystretch}{1.5}
\setlength{\tabcolsep}{4pt} 
\small 
\caption{Marginalized mean values and $1\sigma$ ($68\%$) confidence limits for the model parameters obtained from different dataset combinations. The second column indicates the flat prior ranges used in the MCMC analysis. For parameters with only upper bounds, the limit is indicated. The joint analysis (Combined) represents our baseline constraints.}
\label{tab:constraints}
\begin{tabular}{l c c c c c}
\hline\hline
\textbf{Parameter} & \textbf{Prior Range} & \textbf{CC+SNIa} & \textbf{CC+BAO} & \textbf{SNIa+BAO} & \textbf{Combined} \\
\hline
$H_0$ [km s$^{-1}$ Mpc$^{-1}$]
    & $[60, 80]$
    & $67.9 \pm 1.8$
    & $< 67.9$
    & $67.7^{+2.1}_{-2.5}$
    & $67.8 \pm 1.5$ \\

$\Omega_{m0}$
    & $[0.01, 0.5]$
    & $0.353^{+0.070}_{-0.055}$
    & $0.279^{+0.056}_{-0.034}$
    & $0.252 \pm 0.043$
    & $0.256 \pm 0.039$ \\

$\alpha$ (EGB)
    & $[-0.1, 0.1]$
    & $0.041 \pm 0.033$
    & $-0.003 \pm 0.007$
    & $-0.005 \pm 0.007$
    & $-0.004 \pm 0.007$ \\

$\beta$ (Kaniadakis)
    & $[0.0, 9.8]$
    & $< 5.25$
    & $< 5.33$
    & $2.23^{+0.03}_{-2.20}$
    & $2.26^{+0.11}_{-2.20}$ \\

$\ln c$
    & $[-4.6, 2.3]$
    & $-0.26^{+0.24}_{-0.43}$
    & $0.15^{+0.35}_{-0.54}$
    & $0.19^{+0.28}_{-0.39}$
    & $0.17^{+0.31}_{-0.34}$ \\

$r_{drag}$ [Mpc]
    & $[140, 155]$
    & ---
    & $147.0 \pm 3.0$
    & ---
    & $147.0^{+3.0}_{-3.7}$ \\
\hline\hline
\end{tabular}
\end{table*}

The posteriors in Table~\ref{tab:constraints} lead to the following main conclusions:

\begin{itemize}
    \item \textbf{Consistency with General Relativity:}
    In the joint analysis, the EGB coupling is constrained to $\alpha=-0.004\pm0.007$, which is statistically consistent with $\alpha=0$; thus, current late-time background data do not require a deviation from GR. The constraints nevertheless allow a small negative coupling, corresponding to the GR-connected and theoretically viable branch considered here.

    \item \textbf{Preference for a non-zero $\beta$:}
    The Kaniadakis parameter $\beta$ is only weakly bounded from above by CC and SNIa alone. Once DESI BAO data are included, the posterior favors $\beta\simeq 2.26$ for the full combination. This indicates that allowing a generalized entropy correction can improve the description of the late-time expansion history within this framework.

    \item \textbf{Dark-energy dynamics:}
    The holographic parameter is constrained to $\ln c\simeq0.17$ (i.e., $c\simeq1.18$). In KHDE, the additional $\beta$ term in Eq.~(\ref{eq:wDE_z}) modifies the effective evolution compared to standard HDE, enabling richer dynamics---including phantom-crossing behavior---even when $c>1$.

    \item \textbf{Hubble constant:}
    The inferred value $H_0=67.8\pm1.5~\mathrm{km\,s^{-1}\,Mpc^{-1}}$ is close to the Planck 2018 CMB determination, indicating that this setup favors the lower (early-universe) value of $H_0$ within the considered late-time dataset combination.
\end{itemize}

\begin{figure}[t]
    \centering
    \includegraphics[width=0.8\textwidth]{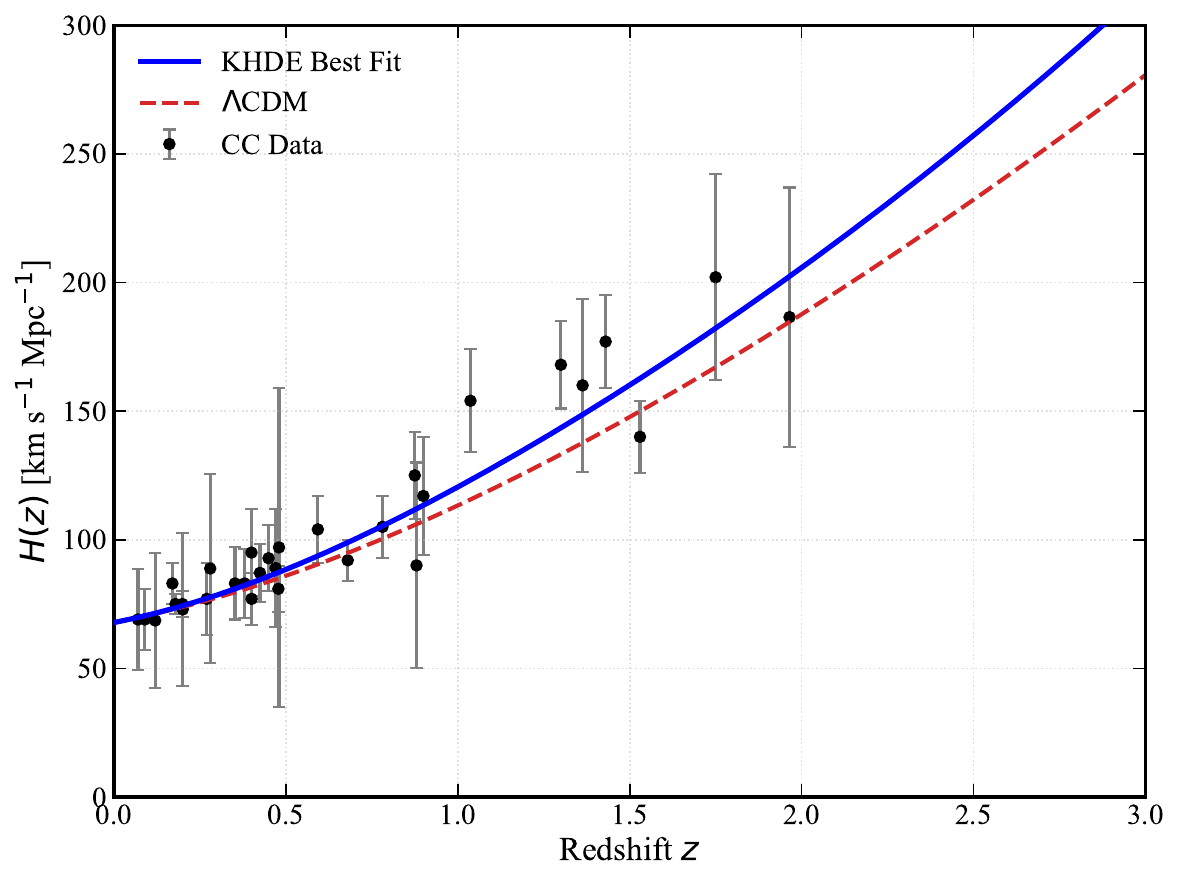}
    \caption{The Hubble parameter $H(z)$ as a function of redshift. The blue solid line represents the best-fit KHDE model ($H_0=67.8, \ln c=0.17, \beta=2.26, \alpha=-0.004$), demonstrating strong agreement with the Cosmic Chronometers data (black dots).}
    \label{fig:hubble_fit}
\end{figure}

For the remainder of this work, we adopt the best-fit values from the full combination,
$(H_0,\alpha,\beta,c)=(67.8,-0.004,2.26,1.18)$,
as our baseline inputs for the black-hole calculations. The agreement between the best-fit expansion history and the CC data is illustrated in Fig.~\ref{fig:hubble_fit}.

\section{EoS Dynamics and Black Hole Mass Evolution}
\label{sec:mass}

With the cosmological parameters constrained in Section~\ref{sec:constraints}, we now use these bounds to model the secular evolution of supermassive black holes. In the standard $\Lambda$CDM paradigm, dark energy is a cosmological constant ($w=-1$), and the Babichev prescription implies no dark-energy-driven mass drift, i.e.~$M(z)\equiv M_0$. In contrast, KHDE predicts a redshift-dependent equation of state $w_{DE}(z)$, which generically induces a non-trivial mass evolution through Eq.~(\ref{eq:mass_acc_z}).

\subsection{Physical Scenarios}

To explore the range of dynamics produced by the holographic parameter $c$ and the Kaniadakis correction $\beta$, we define five representative scenarios consistent with the constraints of Section~\ref{sec:constraints}, fixing $H_0=67.8~\mathrm{km\,s^{-1}\,Mpc^{-1}}$ and $\Omega_{m0}=0.256$. The scenarios are summarized in Table~\ref{tab:scenarios} and are chosen to span the qualitatively distinct late-time behaviors allowed by the parameter space.

For the gravitational sector, we adopt the best-fit EGB coupling $\alpha=-0.004$ as our baseline (solid curves). To isolate the impact of modified gravity, we also show the GR limit $\alpha=0$ (dashed curves). As will be evident below, the differences between EGB and GR predictions are small, indicating that the background dynamics relevant for $w_{DE}(z)$ are largely controlled by the thermodynamic parameters $(c,\beta)$.

\begin{table}[ht]
\centering
\renewcommand{\arraystretch}{1.3}
\setlength{\tabcolsep}{6pt} 
\caption{Definition of the five physical scenarios investigated in this work. The background cosmological parameters are fixed to $H_0 = 67.8$ km s$^{-1}$ Mpc$^{-1}$ and $\Omega_{m0} = 0.256$. The table lists the holographic parameter $c$, the Kaniadakis entropy parameter $\beta$, and the corresponding dynamical regime.}
\label{tab:scenarios}
\begin{tabular}{l c c p{7.5cm}}
\hline \hline
\textbf{Scenario} & \textbf{$c$} & \textbf{$\beta$} & \textbf{Dynamical Regime} \\
\hline
\multicolumn{4}{l}{\textit{\textbf{Reference Model}}} \\
Standard $\Lambda$CDM & -- & -- & Cosmological Constant ($w=-1$) \\
\hline
\multicolumn{4}{l}{\textit{\textbf{KHDE Scenarios}}} \\
Scenario 1 (S1) & $1.18$ & $0.10$ & Non-Monotonic Hump Crossing \\
Scenario 2 (S2) & $1.18$ & $0.35$ & Monotonic Phantom Crossing \\
Scenario 3 (S3) & $1.18$ & $2.26$ & Deep Phantom ($\beta$ at best-fit) \\
Scenario 4 (S4) & $1.18$ & $0.00$ & Standard Quintessence \\
Scenario 5 (S5) & $0.84$ & $0.00$ & Reverse Crossing (Phantom to Quintessence) \\
\hline \hline
\end{tabular}
\end{table}

\subsection{Evolution of the Dark Energy Equation of State}

\begin{figure}[t]
    \centering
    \includegraphics[width=0.8\textwidth]{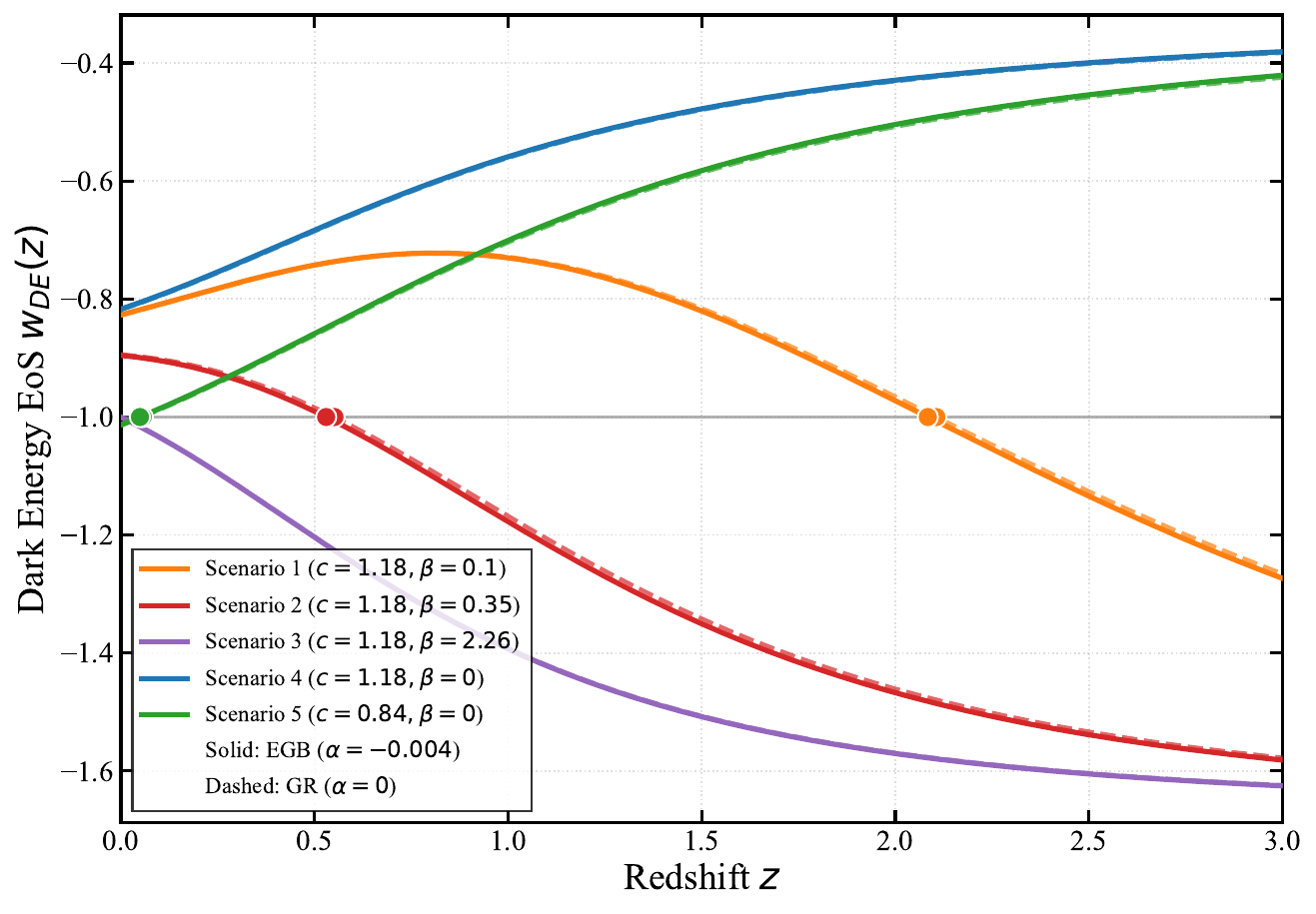}
    \caption{Evolution of the dark energy equation of state $w_{DE}(z)$ for the five physical scenarios defined in Table \ref{tab:scenarios}. Solid lines represent the best-fit EGB gravity ($\alpha=-0.004$), while dashed lines represent General Relativity ($\alpha=0$). The horizontal line marks the Phantom Divide ($w=-1$). The plot highlights diverse dynamical behaviors including monotonic evolution (S3, S4, S5), monotonic crossing (S2), and non-monotonic "hump" crossing (S1). Note the minimal deviation between GR and EGB curves, indicating the perturbative nature of $\alpha$.}
    \label{fig:eos_evolution}
\end{figure}

The accretion rate $\dot M$ is controlled by the factor $(1+w_{DE})$, and thus is highly sensitive to the evolution of the dark-energy equation of state. Figure~\ref{fig:eos_evolution} shows $w_{DE}(z)$ for the scenarios in Table~\ref{tab:scenarios}, illustrating how the interplay between the holographic cutoff and the Kaniadakis correction generates qualitatively different histories.

\textbf{Baseline HDE limit ($\beta=0$):}
We first consider the extensive limit $\beta=0$, for which the dynamics are set solely by $c$. For $c\simeq 1.18>1$ (Scenario~4, blue), the model remains in the quintessence regime with $w>-1$ over the full redshift range shown. In contrast, for $c\simeq 0.84<1$ (Scenario~5, green), the model exhibits a ``Reverse Crossing'': it is phantom-like ($w<-1$) at low redshift and crosses into the quintessence regime at higher redshift.

\textbf{Entropy-induced dynamics ($\beta>0$):}
Turning on Kaniadakis corrections substantially modifies these trajectories. For the best-fit value $\beta=2.26$ (Scenario~3, purple), the non-extensive term dominates and drives the EoS into a ``Deep Phantom'' phase, with $w$ remaining well below $-1$ and evolving monotonically.

More intricate behavior arises when the geometric contribution (set by $c$) and the entropic correction (set by $\beta$) are comparable. A moderate correction $\beta=0.35$ (Scenario~2, red) yields a ``Monotonic Crossing'': the model is quintessence-like at $z=0$ but crosses into the phantom regime as redshift increases. For a smaller correction $\beta=0.10$ (Scenario~1, orange), the EoS develops a non-monotonic ``hump'': it initially increases within the quintessence regime and then turns over, eventually decreasing and crossing the phantom divide. This feature reflects the competition between the evolution of the future event horizon and the entropic modification.

Finally, while varying $\alpha$ produces a small splitting between the solid ($\alpha=-0.004$) and dashed ($\alpha=0$) curves in Fig.~\ref{fig:eos_evolution}, this effect is subdominant relative to the variations induced by $(c,\beta)$.

\subsection{Black Hole Mass Accretion}

\begin{figure}[t]
    \centering
    \includegraphics[width=0.8\textwidth]{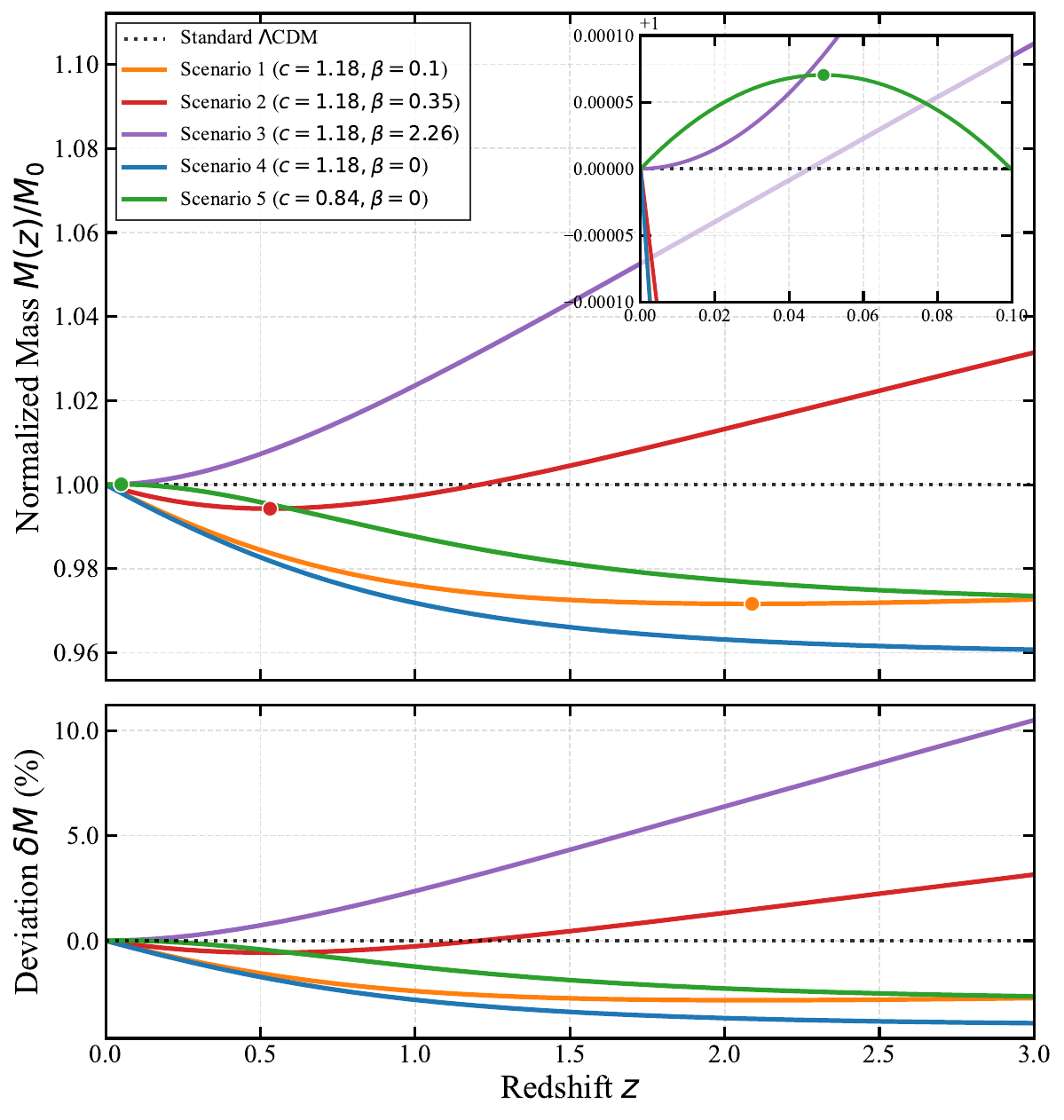}
    \caption{Evolution of the normalized black hole mass $M(z)/M_0$ for the five physical scenarios within the EGB gravity framework ($\alpha=-0.004$). The General Relativity case is omitted due to the negligible impact of $\alpha$ on the background evolution demonstrated earlier. The curves exhibit distinct turning points that strictly correspond to the redshift where $w_{DE}(z)$ crosses $-1$, validating the dynamical link between accretion thermodynamics and the dark energy equation of state.}
    \label{fig:mass_evolution}
\end{figure}

Figure~\ref{fig:mass_evolution} shows the corresponding evolution of the normalized mass $M(z)/M_0$. The most characteristic feature of scenarios with a phantom-divide crossing is the emergence of a turning point in $M(z)/M_0$. This is a direct consequence of Eq.~(\ref{eq:mass_acc_z}): extrema occur precisely when $1+w_{DE}(z)=0$, i.e.~at the crossing redshift, where the accretion term changes sign.

For Scenarios~1 and 2, the mass evolution displays a ``trough'' pattern: $M(z)/M_0$ decreases as one moves to higher redshift, reaches a minimum, and then increases. This reflects a universe that is quintessence-like at $z=0$ ($w>-1$) but becomes phantom-like at higher redshift ($w<-1$), reversing the sign of the secular mass drift. Conversely, Scenario~5 exhibits a ``hump'' pattern, consistent with a transition from a local phantom phase to a higher-redshift quintessence phase.

Beyond these qualitative features, the trajectories show a clear stratification driven primarily by the dark-energy microphysics. Although the EGB coupling $\alpha$ modifies the background expansion through $H(z)$ and hence enters the accretion equation, our EoS analysis indicates that this effect is subdominant. The diversity of accretion histories is instead controlled mainly by the Kaniadakis parameter $\beta$: by reshaping $w_{DE}(z)$, it shifts both the occurrence and direction of phantom crossing and thus determines whether a black hole is in a growth phase (positive accretion) or a mass-loss phase (negative accretion) at a given epoch. This demonstrates that the thermodynamic sector can leave an imprint on the secular evolution of astrophysical black holes within our modeling assumptions.

\section{Redshift Evolution of the Black Hole Shadow}
\label{sec:shadow}

The observable black-hole shadow is set by the critical impact parameter separating photon trajectories that escape to infinity from those captured by the hole. Its characteristic size depends on two ingredients: (i) the intrinsic spacetime geometry, controlled here by the evolving mass $M(z)$ and the EGB coupling $\alpha$, and (ii) propagation effects in the intervening medium. In this section, we first study the shadow evolution in an optical vacuum to isolate the roles of modified gravity and secular mass drift, and then move to a more realistic setting including plasma refraction.

\subsection{Photon Sphere and Shadow Radius in Optical Vacuum}

For null geodesics in the equatorial plane of the 4D EGB spacetime, the effective potential can be written as $V_{\rm eff}(r)=f(r)L^2/r^2$, where $L$ is the photon angular momentum. The radius of the unstable photon sphere $r_{\rm ph}$ is obtained from $V'_{\rm eff}(r_{\rm ph})=0$, which yields
\begin{equation}
    2f(r_{\rm ph})-r_{\rm ph} f'(r_{\rm ph})=0.
    \label{eq:photon_sphere_vacuum}
\end{equation}
In the Schwarzschild limit ($\alpha=0$), one recovers $r_{\rm ph}=3M$, whereas in 4D EGB gravity the solution acquires an $\alpha$ dependence. For an asymptotic observer, the shadow radius is given by the critical impact parameter,
\begin{equation}
    R_{sh}(z)=\xi(\alpha)\,M(z)=\frac{r_{\rm ph}}{\sqrt{f(r_{\rm ph})}}.
\end{equation}
Here, $\xi(\alpha)$ is a dimensionless geometric factor fixed by the EGB coupling, while $M(z)$ encodes the cosmological accretion history.

\subsection{Intrinsic Evolution: The Impact of Phantom Crossing}

\begin{figure}[t]
    \centering
    \includegraphics[width=0.8\textwidth]{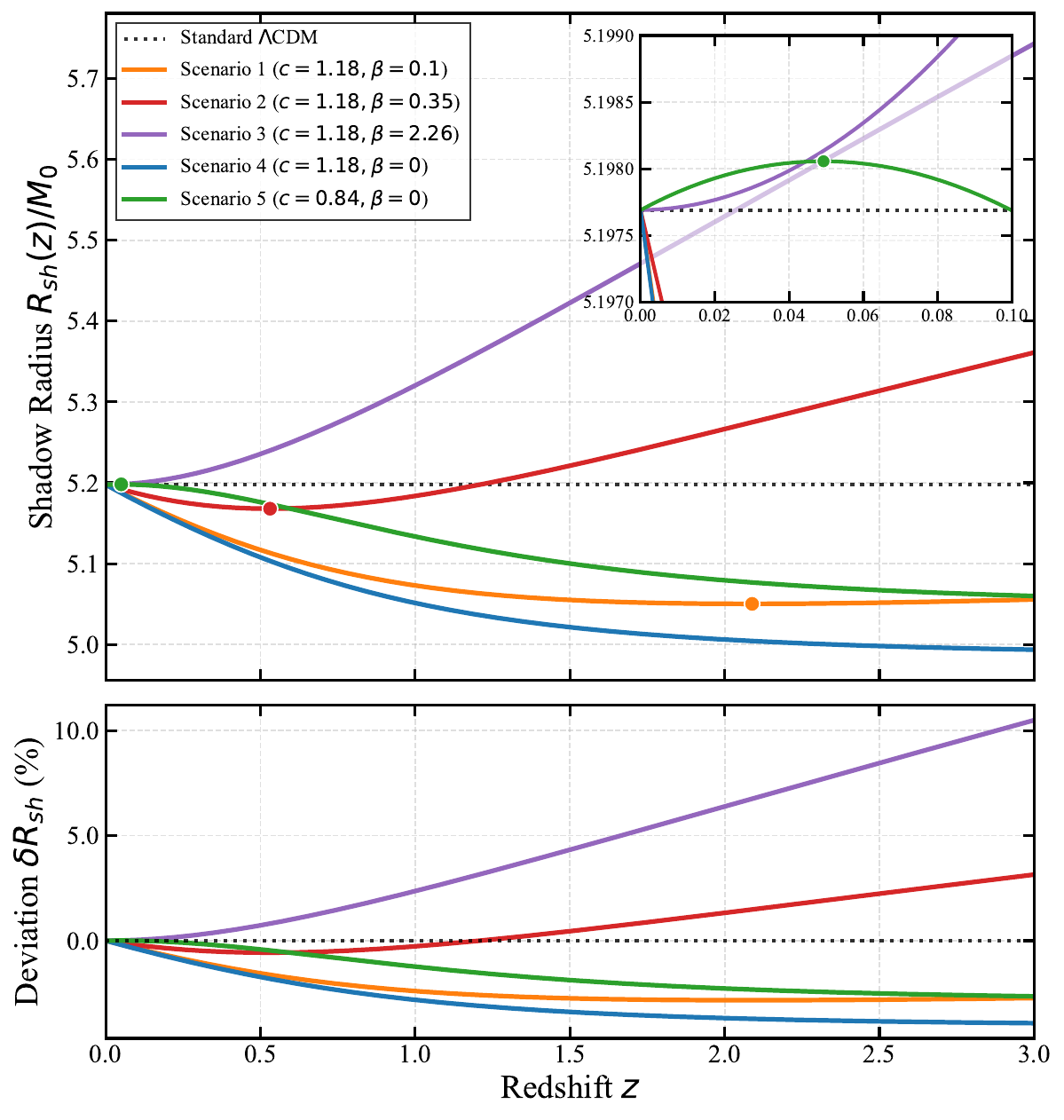}
    \caption{Evolution of the black hole shadow radius $R_{sh}(z)$ in an optical vacuum ($n=1$) for the five physical scenarios under EGB gravity ($\alpha=-0.004$). The evolution morphology is primarily dictated by the mass accretion history $M(z)$, exhibiting the same non-monotonic features ("hump" or "trough" shapes) driven by the phantom divide crossing. The absolute scale is modulated by the geometric factor $\xi(\alpha)$.}
    \label{fig:shadow_vacuum}
\end{figure}

To disentangle intrinsic effects from environmental ones, we first set $n=1$ (optical vacuum). Figure~\ref{fig:shadow_vacuum} shows the evolution of $R_{sh}(z)$ (in units of the reference mass $M_0$) for the five scenarios in the EGB case $\alpha=-0.004$.

Because $R_{sh}(z)$ scales with $M(z)$ in vacuum, the shadow evolution directly traces the secular mass history discussed in Section~\ref{sec:mass}. In particular, Scenarios~1 and 2 exhibit a ``trough'': $R_{sh}$ decreases at low redshift and then turns around and increases at higher redshift, mirroring the change in the sign of $1+w_{DE}(z)$ across the phantom divide. Conversely, Scenario~5 shows a ``hump'' behavior, corresponding to the opposite transition (phantom at low redshift to quintessence at higher redshift).

While the \emph{shape} of the curves is driven by the thermodynamic parameters $(c,\beta)$ through the accretion history, the \emph{overall normalization} is set by the EGB coupling via $\xi(\alpha)$. For our best-fit value $\alpha=-0.004$, $\xi(\alpha)$ is slightly larger than its GR counterpart ($3\sqrt{3}\simeq 5.196$), implying a nearly redshift-independent enhancement of the shadow size relative to GR. This constant geometric offset is distinct from the secular evolution induced by $M(z)$, and therefore helps separate the modified-gravity contribution (baseline scale) from the dark-energy contribution (evolution and turning points).

In summary, the vacuum shadow evolution encodes a twofold signature: $\alpha$ sets the overall scale through geometry, whereas $(c,\beta)$ governs the redshift dependence through the KHDE-driven mass drift and any associated turning points at phantom crossing.

\subsection{Impact of Plasma Environment}
\label{sec:plasma}

In realistic settings, photon propagation occurs through a dispersive plasma. The plasma induces a frequency-dependent refractive index $n$, modifying photon trajectories and thereby the apparent shadow size. In a cosmological context, one may further allow for redshift trends in the effective plasma properties.

We model the medium as a cold, non-magnetized dispersive plasma. For a photon with local angular frequency $\omega(r)$, the refractive index satisfies the dispersion relation \cite{Perlick2015}
\begin{equation}
    n^2 = 1 - \frac{\omega_p(r)^2}{\omega(r)^2},
\end{equation}
where $\omega_p(r)$ is the electron plasma frequency,
$\omega_p^2 = 4\pi e^2 N_e(r)/m_e$,
with $e$ the electron charge, $m_e$ the electron mass, and $N_e(r)$ the electron number density. Since $n<1$, the plasma acts as a dispersive medium and refracts light rays, altering the critical impact parameter \cite{Perlick2015, Xu:2025iwg}.

To connect local propagation to global evolution, we incorporate two competing redshift-dependent effects through a \textbf{cosmological scaling ansatz} intended as a statistical baseline:

\begin{enumerate}
    \item \textbf{Density Evolution ($\omega_p$ term):} We assume that the characteristic plasma density of the cosmic population scales with the expansion history. Although local astrophysical processes may decouple individual sources from the Hubble flow, the average density of the intergalactic medium feeding these systems evolves as $\rho \propto (1+z)^3$. As a statistical baseline, we impose a similar scaling on the plasma number density: $N_e(z) \propto (1+z)^3$. Consequently, $\omega_p^2(z) \propto (1+z)^3$.
    \item \textbf{Frequency Redshift ($\omega$ term):} Due to the cosmic expansion, the photon frequency $\omega$ observed at a high redshift $z$ relates to the observed frequency $\omega_0$ at $z=0$ via the standard relation $\omega(z) = \omega_0(1+z)$. Squaring this gives $\omega^2(z) \propto (1+z)^2$.
\end{enumerate}

Combining these scalings implies
\begin{equation}
    \frac{\omega_p^2(z)}{\omega^2(z)} \propto \frac{(1+z)^3}{(1+z)^2}=(1+z).
\end{equation}
Assuming in addition a power-law radial profile $N_e(r)\propto r^{-h}$, we adopt the effective refractive-index model
\begin{equation}
    n(r,z)=\sqrt{1-\frac{k_0(1+z)}{r^h}},
    \label{eq:refractive_index}
\end{equation}
where $h$ sets the radial fall-off and $k_0$ is a dimensionless normalization that controls the plasma strength at $z=0$. Equation~(\ref{eq:refractive_index}) should be viewed as a toy model that defines a baseline redshift trend; individual systems may deviate from it, but population-level analyses could in principle probe the associated systematic evolution.

\begin{figure}[t]
    \centering
    \begin{subfigure}[b]{0.48\textwidth}
        \centering
        \includegraphics[width=\textwidth]{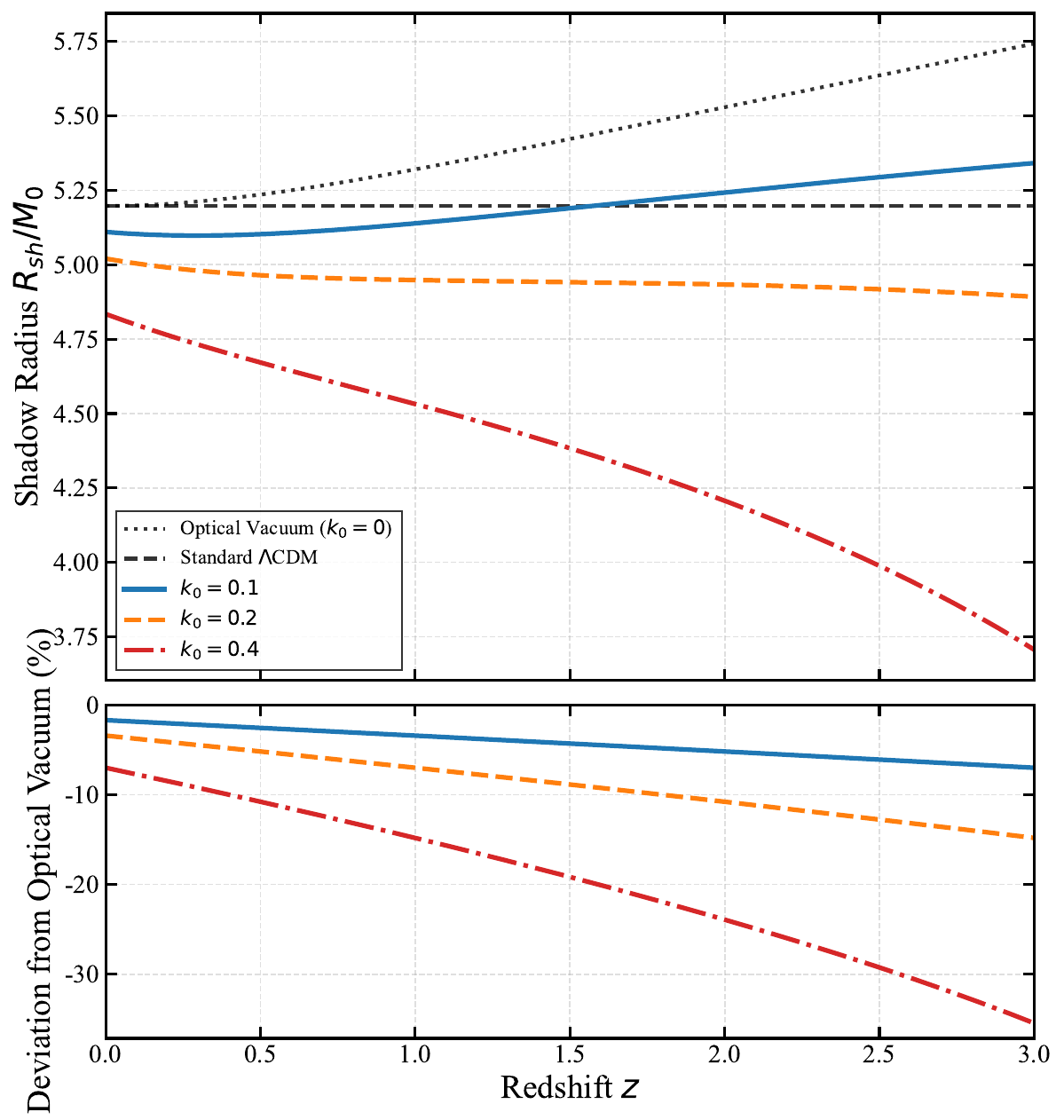}
        \caption{Sensitivity to plasma density $k_0$}
        \label{fig:plasma_a}
    \end{subfigure}
    \hfill
    \begin{subfigure}[b]{0.48\textwidth}
        \centering
        \includegraphics[width=\textwidth]{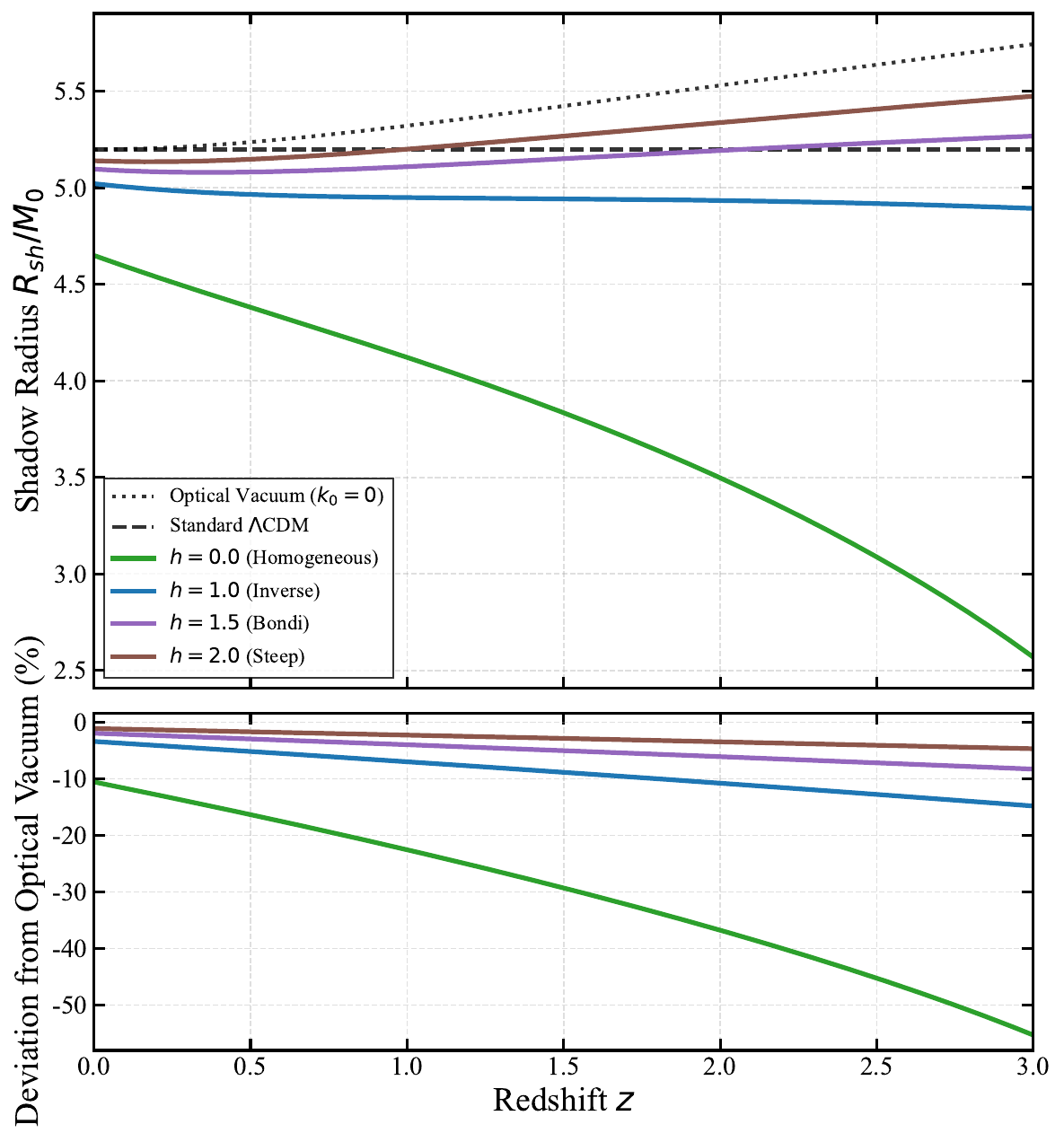}
        \caption{Sensitivity to radial profile $h$}
        \label{fig:plasma_b}
    \end{subfigure}
    \caption{\textbf{Impact of Plasma Environment.} (a) Comparison of shadow evolution in optical vacuum ($k_0=0$) versus dispersive plasma ($k_0>0$) for fixed $h=1$. Increasing $k_0$ progressively suppresses the shadow size at high redshift. (b) Dependence on the radial profile slope $h$ for fixed $k_0=0.2$. Larger $h$ yields larger shadows (closer to the vacuum limit) because the density drops more rapidly near the photon sphere. The baseline model corresponds to the best-fit mean values ($c=1.18, \beta=2.26, \alpha=-0.004$).}
    \label{fig:plasma_sensitivity}
\end{figure}

\subsubsection{The Dominance of Environmental Evolution}

In the presence of plasma, the observable shadow radius becomes
\begin{equation}
    R_{sh}(z)=\frac{r_{\rm ph}\,n(r_{\rm ph},z)}{\sqrt{f(r_{\rm ph})}}.
\end{equation}
Because the plasma term $k_0(1+z)/r^h$ in Eq.~(\ref{eq:refractive_index}) grows with redshift, the refractive index $n$ decreases, and the plasma acts as an increasingly strong dispersive screen at earlier epochs.

This competition between refractive suppression and intrinsic mass evolution is illustrated in Fig.~\ref{fig:plasma_sensitivity}(a), where we compare the vacuum case ($k_0=0$, black dotted curve) to increasing plasma strengths ($k_0=0.1,0.2,0.4$) at fixed $h=1$. As $k_0$ increases, the intrinsic trend is progressively damped and can be effectively reversed. For $k_0=0.2$, the shadow at $z=3$ is substantially suppressed relative to the vacuum prediction. This environmental screening can therefore mask the mass-drift signal, implying that sufficiently dense plasma environments may yield smaller apparent shadows at higher redshift even when the intrinsic (vacuum) shadow would increase.

\subsubsection{Sensitivity to Accretion Profile}

We next test the dependence on the radial profile parameter $h$ in Fig.~\ref{fig:plasma_sensitivity}(b), keeping $k_0=0.2$ fixed. The parameter $h$ controls how rapidly the plasma density decreases with radius. Larger $h$ implies lower densities near the photon sphere ($r_{\rm ph}\approx 3M$) and hence $n\to1$, driving the shadow closer to the vacuum limit (e.g.~the $h=2.0$ curve). Conversely, smaller $h$ maintains higher densities near the photon sphere, producing stronger refraction and a smaller shadow. Across the explored range, the qualitative outcome is robust: plasma introduces a systematic reduction in the observable shadow size that becomes more pronounced with increasing redshift. This indicates that while the local profile (via $h$) modulates the overall magnitude, the redshift trend driven by the $(1+z)$ factor is the primary source of the evolving screening effect.

\section{Conclusion and Discussion}
\label{sec:conclusion}

In this work, we constructed a unified framework connecting global cosmological expansion to local strong-field observables. By embedding 4D Einstein--Gauss--Bonnet (EGB) black holes in a Kaniadakis holographic dark energy (KHDE) cosmology, we examined how modified gravity, dynamical dark-energy accretion, and dispersive plasma jointly affect the secular evolution and observable shadow properties of supermassive black holes.

We first constrained the background cosmology through an MCMC analysis using Cosmic Chronometers (CC), Type Ia supernovae (SNIa), and DESI BAO data. The combined fit favors a holographic parameter $c\simeq 1.18$ and a Kaniadakis parameter $\beta\simeq 2.26$. Although the posterior mean points to a phantom-like regime driven by the non-extensive sector, the constraint on $\beta$ remains broad ($2.26^{+0.11}_{-2.20}$), so that the extensive holographic limit $\beta\to 0$ is still consistent with the data at $1\sigma$. In the gravity sector, we find $\alpha=-0.004\pm 0.007$, indicating that any departure from GR at the background level is, at most, perturbative within current late-time constraints.

Using these bounds, we then modeled the secular mass drift induced by KHDE accretion. Treating the best-fit values as a representative baseline, we showed that $\beta$ acts as a dynamical control parameter: while the standard holographic limit ($\beta=0$) yields monotonic mass evolution, non-extensive corrections can generate non-monotonic ``hump'' or ``trough'' patterns, with extrema tracking the phantom-divide crossing epoch ($w_{DE}=-1$). In the deep-phantom scenario, the accretion term becomes negative, implying a net decrease of the black-hole mass with cosmic time within the adopted Babichev prescription. We also found that varying $\alpha$ primarily rescales the shadow through a geometric factor, providing a largely redshift-independent offset that helps separate spacetime-geometry effects from the thermodynamically driven mass evolution.

Finally, we assessed how these intrinsic signatures map to observable shadows once plasma propagation effects are included. Adopting a cosmological scaling ansatz for the plasma term (with $\rho_{\rm plasma}\propto (1+z)^3$), we found a strong competition between intrinsic evolution and environmental screening: although phantom-like dynamics can enlarge the vacuum shadow at higher redshift, increasing plasma refraction tends to suppress the observed shadow and can dominate the net trend. For representative plasma strengths (e.g.~$k_0\sim 0.2$), the environmental effect overwhelms the intrinsic signal at high redshift, leading to an apparent monotonic shrinkage relative to the vacuum expectation.

Overall, our results emphasize both a challenge and an opportunity. The dominant redshift trend of the shadow can be largely set by the evolving medium and hence be highly degenerate. Nevertheless, departures from the refractive baseline---including the redshift dependence and any residual turning-point structure inherited from $w_{DE}(z)$---encode information about the underlying dark-energy dynamics. Future horizon-scale observations, especially if combined through population-level stacking across redshift bins to reduce source-to-source environmental scatter, may therefore provide a complementary avenue to probe the thermodynamic sector of dark energy in the strong-gravity regime.

\section*{Acknowledgments}
This work was supported by the National Natural Science Foundation of China under Grant No. 12305070, and the Basic Research Program of Shanxi Province under Grant Nos. 202303021222018 and 202303021221033.
\bibliographystyle{unsrt}

\bibliography{references}

\end{document}